\documentclass[12pt]{iopart}

\usepackage{graphicx,setspace}
\usepackage{dsfont}
\usepackage{amssymb}
\usepackage{mathrsfs}
\usepackage{psfrag}

\begin{document}
\title{Casimir effect from macroscopic quantum electrodynamics}
\author{T G Philbin}
\address{School of Physics and Astronomy, University of St Andrews,
North Haugh, St Andrews, Fife KY16 9SS,
Scotland, UK.}
\ead{tgp3@st-andrews.ac.uk}

\begin{abstract}
The canonical quantization of macroscopic electromagnetism was recently presented in {\it New J.\ Phys.}\ {\bf 12} (2010) 123008. This theory is here used to derive the Casimir effect, by considering the special case of thermal and zero-point fields. The stress-energy-momentum tensor of the canonical theory follows from Noether's theorem, and its electromagnetic part in thermal equilibrium gives the Casimir energy density and stress tensor. The results hold for arbitrary inhomogeneous magnetodielectrics and are obtained from a rigorous quantization of electromagnetism in dispersive, dissipative media. Continuing doubts about the status of the standard Lifshitz theory as a proper quantum treatment of Casimir forces do not apply to the derivation given here. Moreover, the correct expressions for the Casimir energy density and stress tensor inside media follow automatically from the simple restriction to thermal equilibrium, without the need for complicated thermodynamical or mechanical arguments.
\end{abstract}


\section{Introduction}
The phenomenon of forces on macroscopic objects due to electromagnetic zero-point and thermal fields is usually called the Casimir effect, after the first, highly idealized, theoretical result in the subject~\cite{cas48}. A theory of the effect for realistic materials was given by Lifshitz and co-workers~\cite{lif55, dzy61, LL} and this remains the most general formalism for describing the local electromagnetic quantities that produce the forces. Some authors prefer other terminology, such as Casimir-Lifshitz effect, or van der Waals forces, to designate this same phenomenon; in this paper the term Casimir effect is used, both for the zero-point and thermal contributions.

Despite its good agreement with experiments that measure the zero-point Casimir force, there have been been persistent doubts about the status of Lifshitz theory as a quantum theory. These doubts are voiced, for example, in two recent publications~\cite{bar10,ros10}, and a detailed analysis in~\cite{ros10} concludes that ``Lifshitz theory is actually a classical stochastic electrodynamical theory''. The source of these doubts is not difficult to find if one examines the details of Lifshitz theory. As the Casimir effect is a phenomenon of quantum electromagnetism in the presence of macroscopic media, one would expect the general theory of the effect to be based on the principles of quantum electrodynamics. Lifshitz theory, however, is based rather on the the principles of thermodynamics. In fact there is no Hamiltonian in Lifshitz theory and there are no quantized fields, yet the formalism seeks to calculate the forces caused by quantum zero-point fields (as well as by thermal fields).

The requirements of a quantum theory of light in media that can serve as a basis for the Casimir effect (among many other phenomena) are easy to identify. The theory must hold for arbitrary magnetodielectrics in order to be applicable to the realistic materials used in experiments, materials whose optical properties (dielectric functions) are known only through measurement. As the Casimir effect is a broadband phenomenon in which all frequencies must be included, the theory must also take full account of material dispersion and absorption. The classical theory of light in these circumstances is of course the macroscopic Maxwell equations, where the electromagnetic properties of the media are encompassed in arbitrary electric permittivities and magnetic permeabilities obeying the Kramers-Kronig relations. The theoretical basis of the Casimir effect should therefore be a quantum theory of Maxwell's macroscopic electromagnetism. But whereas the quantum theory of the free-space Maxwell equations, quantum electrodynamics (QED), was the foundation of quantum field theory, and led to a general formalism for quantizing classical field theories, this quantization procedure was not applied to macroscopic electromagnetism. What has become known in the literature as macroscopic QED is not based on the rules for quantizing field theories, but is instead a phenomenological theory wherein no rigorous quantization is attempted (see~\cite{kno01,sch08} for detailed presentations). This phenomenological procedure is subject to much of the criticism directed at Lifshitz theory, and it will be seen that the results derived in this paper require, among other things, an action principle, something that is lacking in the phenomenological approach. The quantization rules of quantum field theory were not obviously applicable in the case of macroscopic electromagnetism due to the complications of dispersion and dissipation, and this is what led to the phenomenological approach. In fact it had been ``widely agreed"~\cite{hut92} that a proper quantization of macroscopic electromagnetism could not be performed, and that only in cases where a simple microscopic model of the dielectric functions of the medium is explicitly introduced could the standard quantization rules be applied. In~\cite{phi10}, however, the canonical quantization of macroscopic electromagnetism was achieved, providing a rigorous macroscopic QED and removing the need for a phenomenological approach. Since the macroscopic QED derived in~\cite{phi10} applies to arbitrary magnetodielectrics and takes full account of dispersion and absorption, it meets the criteria outlined above for a rigorous quantum foundation for the Casimir effect. This paper derives the Casimir effect from macroscopic QED by considering the special case of thermal equilibrium.\footnote{Hereafter macroscopic QED refers to the canonically quantized theory presented in~\cite{phi10}, not the previous phenomenological formalism~\cite{kno01,sch08}.}

A brief summary of macroscopic QED is given in section~\ref{sec:mQED}. The stress-energy-momentum tensor of macroscopic electromagnetism is derived in full generality in section~\ref{sec:em}, by application of Noether's theorem to the action principle given in~\cite{phi10}. Specialization to thermal equilibrium (including the zero-point fields) is made in section~\ref{sec:thermal}. The expectation value of the electromagnetic part of the stress-energy-momentum tensor in thermal equilibrium gives the Casimir effect. Correlation functions of the quantum field operators in thermal equilibrium are calculated in~\ref{sec:cor} and used in sections~\ref{sec:en} and~\ref{sec:st} to obtain the Casimir energy density and stress tensor.

\section{Macroscopic QED}  \label{sec:mQED}
This section summarizes the results of macroscopic QED~\cite{phi10} that we require to derive the Casimir effect. The action of macroscopic electromagnetism is~\cite{phi10}
\begin{equation} \label{S}
\fl
S[\phi,\mathbf{A},\mathbf{X}_\omega,\mathbf{Y}_\omega]=S_{\mathrm{em}}[\phi,\mathbf{A}]+S_ \mathrm {X}[\mathbf{X}_\omega]+S_ \mathrm{Y}[\mathbf{Y}_\omega]+S_{\mathrm{int}}[\phi,\mathbf{A},\mathbf{X}_\omega,\mathbf{Y}_\omega],
\end{equation}
where $S_{\mathrm{em}}$ is the free electromagnetic action
\begin{equation} \label{Sem}
S_{\mathrm{em}}[\phi,\mathbf{A}]=\frac{\kappa_0}{2}\int\rmd^4 x\left(\frac{1}{c^2}\mathbf{E}\cdot\mathbf{E}-\mathbf{B}\cdot\mathbf{B}\right), \quad \kappa_0=1/\mu_0,
\end{equation}
$S_ \mathrm{X}$ and $S_ \mathrm{Y}$ are the actions for free reservoir oscillators:
\begin{eqnarray}
S_ \mathrm{X}[\mathbf{X}_\omega]=\frac{1}{2}\int\rmd^4 x\int_0^\infty\rmd\omega\left(\partial_t\mathbf{X}_\omega\cdot\partial_t\mathbf{X}_\omega-\omega^2\mathbf{X}_\omega\cdot\mathbf{X}_\omega\right),  \\[3pt]
S_ \mathrm{Y}[\mathbf{Y}_\omega]=\frac{1}{2}\int\rmd^4 x\int_0^\infty\rmd\omega\left(\partial_t\mathbf{Y}_\omega\cdot\partial_t\mathbf{Y}_\omega-\omega^2\mathbf{Y}_\omega\cdot\mathbf{Y}_\omega\right),
\end{eqnarray}
and $S_{\mathrm{int}}$ is the interaction part of the action, coupling the electromagnetic fields to the reservoir:
\begin{eqnarray}
S_{\mathrm{int}}[\phi,\mathbf{A},\mathbf{X}_\omega,\mathbf{Y}_\omega]=\int\rmd^4 x\int_0^\infty\rmd\omega\left[\alpha(\mathbf{r},\omega)\mathbf{X}_\omega\cdot\mathbf{E}+\beta(\mathbf{r},\omega)\mathbf{Y}_\omega\cdot\mathbf{B}\right],  \label{Sint} \\[3pt]
\alpha(\mathbf{r},\omega)=\left[\frac{2\varepsilon_0}{\pi}\omega\varepsilon_\mathrm{I}(\mathbf{r},\omega)\right]^{1/2}, \qquad \beta(\mathbf{r},\omega)=\left[-\frac{2\kappa_0}{\pi}\omega\kappa_\mathrm{I}(\mathbf{r},\omega)\right]^{1/2}.  \label{ab}
\end{eqnarray}
The imaginary parts of the dielectric functions of the medium appear in the coupling functions (\ref{ab}); their real parts are given by the Kramers-Kronig relation~\cite{LLcm,jac}:
\begin{equation}  \label{KK}
\varepsilon_\mathrm{R}(\mathbf{r},\omega')-1=\frac{2}{\pi}\mathrm{P}\int_0^\infty\rmd\omega\frac{\omega\varepsilon_\mathrm{I}(\mathbf{r},\omega)}{\omega^2-\omega'^2},  \qquad \mbox{and similarly for $\kappa(\mathbf{r},\omega)$.}
\end{equation}
As in~\cite{phi10}, we assume the medium is isotropic, with scalar dielectric functions $\varepsilon(\mathbf{r},\omega)$ and $\kappa(\mathbf{r},\omega)=1/\mu (\mathbf{r},\omega)$; anisotropy can be included by obvious modifications. The field equations of the action (\ref{S})--(\ref{ab}) are 
\begin{eqnarray}
\varepsilon_0\nabla\cdot\mathbf{E}+\int_0^\infty\rmd\omega\,\nabla\cdot\left[\alpha(\mathbf{r},\omega)\mathbf{X}_\omega\right]=0,   \label{Eeq} \\[3pt]
-\kappa_0\nabla\times\mathbf{B}+\varepsilon_0\partial_t\mathbf{E}+\int_0^\infty\rmd\omega\left\{\alpha(\mathbf{r},\omega) \partial_t\mathbf{X}_\omega+\nabla\times\left[\beta(\mathbf{r},\omega)\mathbf{Y}_\omega\right]\right\}=0,  \label{Beq} \\[3pt]
-\partial_t^2\mathbf{X}_\omega-\omega^2\mathbf{X}_\omega+\alpha(\mathbf{r},\omega)\mathbf{E}=0,  \label{Xeq}  \\
-\partial_t^2\mathbf{Y}_\omega-\omega^2\mathbf{Y}_\omega+\beta(\mathbf{r},\omega)\mathbf{B}=0.  \label{Yeq}
\end{eqnarray}
By solving the equations for the reservoir fields $\mathbf{X}_\omega$ and $\mathbf{Y}_\omega$, the independent equations of the electromagnetic fields are found and are precisely the macroscopic Maxwell equations~\cite{phi10}
\begin{eqnarray}
\nabla\cdot\mathbf{D}=\sigma, \label{gauss} \\
\nabla\times\mathbf{H}-\partial_t\mathbf{D}=\mathbf{j},  \label{amp}
\end{eqnarray}
where the charge density $\sigma$ and current density $\mathbf{j}$ are given in terms of arbitrary free-field solutions for the reservoir fields $\mathbf{X}_\omega$ and $\mathbf{Y}_\omega$. The other two Maxwell equations are identities in terms of the potentials $\mathbf{A}$ and $\phi$. 

As the action (\ref{S})--(\ref{ab}) features the dynamical fields and their first derivatives, canonical quantization can proceed without difficulty~\cite{phi10}. The resulting Hamiltonian can be diagonalized to the form
\begin{equation} \label{Hdiag}
\hat{H}=\sum_{\lambda=\mathrm{e},\mathrm{m}}\int\rmd^3 \mathbf{r}\int_0^\infty\rmd\omega\,\hbar\omega\mathbf{\hat{C}}^\dagger_\lambda(\mathbf{r},\omega)\cdot\mathbf{\hat{C}}_\lambda (\mathbf{r},\omega),
\end{equation}
where the diagonalizing eigenmode creation and annihilation operators obey
\begin{equation} \label{C,C}
\fl
\left[\hat{C}_{\lambda i}(\mathbf{r},\omega),\hat{C}^\dagger_{\lambda' j}(\mathbf{r'},\omega')\right]=\delta_{ij}\delta_{\lambda\lambda'} \delta(\omega-\omega') \delta(\mathbf{r}-\mathbf{r'}), \qquad \left[\hat{C}_{\lambda i}(\mathbf{r},\omega),\hat{C}_{\lambda' j}(\mathbf{r'},\omega')\right]=0.
\end{equation}
The infinite zero-point energy has been omitted from the Hamiltonian (\ref{Hdiag}). Zero-point energy is of course crucial for the Casimir effect, but the electromagnetic energy density and stress tensor will be calculated below taking full account of the zero-point fields. The quantum macroscopic Maxwell equations 
\begin{eqnarray}
\nabla\cdot\mathbf{\hat{D}}= \hat{\sigma}, \label{qgauss}  \\
\nabla\times\mathbf{\hat{H}}-\partial_t\mathbf{\hat{D}}=\mathbf{\hat{j}},  \label{qamp}
\end{eqnarray}
hold with the charge and current density operators given, in the frequency domain, by
\begin{eqnarray}
\fl
\hat{\sigma} (\mathbf{r}, \omega)=\frac{1}{\rmi\omega}\nabla \cdot\mathbf{\hat{j}}(\mathbf{r}, \omega) =-2\pi\nabla\cdot\left\{\left[\frac{\hbar\varepsilon_0}{\pi}\varepsilon_\mathrm{I}(\mathbf{r},\omega)\right]^{1/2}\mathbf{\hat{C}}_\mathrm{e}(\mathbf{r},\omega)\right\},   \label{qsigma} \\[3pt]
\fl
\mathbf{\hat{j}} (\mathbf{r}, \omega)=-2\pi\rmi \omega\left[\frac{\hbar\varepsilon_0}{\pi}\varepsilon_\mathrm{I}(\mathbf{r},\omega)\right]^{1/2}\mathbf{\hat{C}}_\mathrm{e}(\mathbf{r},\omega) 
+2\pi\nabla\times\left\{\left[-\frac{\hbar\kappa_0}{\pi}\kappa_\mathrm{I}(\mathbf{r},\omega)\right]^{1/2}\mathbf{\hat{C}}_\mathrm{m}(\mathbf{r},\omega)\right\}. \label{jopdef}
\end{eqnarray}
The relationship between fields in the time and frequency domains is, for the example of the electric field,
\begin{equation}  \label{Efreq}
\mathbf{\hat{E}}(\mathbf{r},t)=\frac{1}{2\pi}\int_0^\infty \rmd\omega\left[\mathbf{\hat{E}} (\mathbf{r},\omega)\exp(-\rmi\omega t)+\mbox{c.c.}\right].
\end{equation}
The electromagnetic field operators and the reservoir field operators are also expressed in terms of the diagonalizing operators, as follows. For the electric field operator the relation follows from (\ref{jopdef}) and
\begin{equation}
\fl
\mathbf{\hat{E}}(\mathbf{r},t)=\frac{\mu_0}{2\pi}\int_0^\infty\rmd \omega\int\rmd^3\mathbf{r'}\left[\rmi \omega \mathbf{G}(\mathbf{r},\mathbf{r'},\omega)\cdot\mathbf{\hat{j}} (\mathbf{r'}, \omega)\exp(-\rmi \omega t)+\mbox{h.c.}\right],  \label{EopG} 
\end{equation}
where the Green bi-tensor $\mathbf{G}(\mathbf{r},\mathbf{r'},\omega)$ is the solution of
\begin{equation} \label{green}
\nabla\times\left[\kappa(\mathbf{r},\omega)\nabla\times\mathbf{G}(\mathbf{r},\mathbf{r'}, \omega)\right]-\frac{\omega^2}{c^2}\varepsilon(\mathbf{r},\omega)\mathbf{G}(\mathbf{r},\mathbf{r'}, \omega)=\mathds{1}\delta(\mathbf{r}-\mathbf{r'}).
\end{equation}
The magnetic field operator is 
\begin{equation} \label{BE}
\mathbf{\hat{B}}(\mathbf{r},\omega)=-\rmi\nabla\times\mathbf{\hat{E}}(\mathbf{r},\omega)/\omega.
\end{equation}
Finally, the reservoir field operators $\mathbf{\hat{X}}_\omega$ and $\mathbf{\hat{Y}}_\omega$ can be written in the frequency domain as
\begin{eqnarray}
\mathbf{\hat{X}}_\omega(\mathbf{r},\omega')=&2\pi\sqrt{\frac{\hbar}{2\omega}}\delta(\omega-\omega') \mathbf{\hat{C}}_\mathrm{e}(\mathbf{r},\omega')  \nonumber \\[3pt]
& +\frac{\alpha(\mathbf{r},\omega)}{2\omega}\left[\frac{1}{\omega-\omega'-\rmi0^+}+\frac{1}{\omega+\omega'}\right]\mathbf{\hat{E}}(\mathbf{r},\omega'),   \label{XCE}  \\[3pt]
\mathbf{\hat{Y}}_\omega(\mathbf{r},\omega')=& 2\pi\sqrt{\frac{\hbar}{2\omega}}\delta(\omega-\omega') \mathbf{\hat{C}}_\mathrm{m}(\mathbf{r},\omega')   \nonumber \\[3pt]
&-\frac{\rmi\beta(\mathbf{r},\omega)}{2\omega'\omega}\left[\frac{1}{\omega-\omega'-\rmi0^+}+\frac{1}{\omega+\omega'}\right]\nabla\times\mathbf{\hat{E}}(\mathbf{r},\omega').  \label{YCB} 
\end{eqnarray}  
The last two equations were not explicitly written in~\cite{phi10}; they are obtained from the expansion (58) in~\cite{phi10} of the operator $\mathbf{\hat{X}}_\omega$ in terms of the diagonalizing operators, and the analogous expansion of $\mathbf{\hat{Y}}_\omega$, by inserting the results (70), (71), (76) and (88) of~\cite{phi10}. Use is also made of the fact that all frequency arguments are taken only at positive values (see (\ref{Efreq})), so that the sums of frequencies in denominators do not give rise to poles.

\section{Stress-energy-momentum tensor of macroscopic electromagnetism} \label{sec:em}
The canonical formulation of macroscopic electromagnetism given in~\cite{phi10} leads directly to the stress-energy-momentum tensor of the system, in both the classical and quantum cases. As dissipation of electromagnetic energy by the medium is fully taken into account through the presence of the reservoir fields, a conserved total energy of the system exists (if the dielectric functions of the medium are time independent). Similarly, total momentum is conserved if the medium is homogeneous.

The complete stress-energy-momentum tensor of the system follows from application of Noether's theorem to the action (\ref{S})--(\ref{ab}). We state the results for classical fields, but the quantum stress-energy-momentum tensor is of the same form because the quantum field operators obey the same dynamical equations as the classical fields, as shown in~\cite{phi10}. 

\subsection{Energy density and energy flux}
The invariance of the action (\ref{S})--(\ref{ab}) under active time translations of the dynamical fields implies a conservation law that can be extracted as follows~\cite{wei}. We make an active infinitesimal time translation of all the dynamical fields---in the case of the vector potential, $\mathbf{A}(\mathbf{r},t)\rightarrow\mathbf{A}(\mathbf{r},t+\zeta(\mathbf{r},t))$---but take the translation $\zeta(\mathbf{r},t)$ to vary in space and time. The resulting change in the action can be reduced to the form
\begin{equation} \label{actvart}
\delta S=\int d^4x\left(\rho\,\partial_t\zeta+\mathbf{s}\cdot\nabla\zeta\right),
\end{equation}
where $\rho$ is the energy density and $\mathbf{s}$ is the energy flux, obeying the conservation law
\begin{equation} \label{cone}
\partial_t \rho+ \nabla \cdot\mathbf{s}=0.
\end{equation}
The calculation is straightforward and yields
\begin{eqnarray}
\fl
\rho=&\frac{\kappa_0}{2}\left[\frac{1}{c^2}\mathbf{E}\cdot(-\partial_t\mathbf{A}+\nabla\phi)+\mathbf{B}^2\right] \nonumber \\[3pt]
\fl
&+\int_0^\infty\rmd\omega\left[\frac{1}{2}(\partial_t\mathbf{X}_\omega)^2+\frac{1}{2}(\partial_t\mathbf{Y}_\omega)^2+\frac{1}{2}\omega^2(\mathbf{X}_\omega^2+\mathbf{Y}_\omega^2)+\alpha\mathbf{X}_\omega\cdot\nabla\phi-\beta\mathbf{Y}_\omega\cdot\mathbf{B}\right], \label{rho1} \\[3pt]
\fl
s^i=&-\kappa_0\left[\frac{1}{c^2}E^i\partial_t\phi+2\nabla^{[i}A^{j]}\partial_tA_j\right]-\int_0^\infty\rmd\omega\left[\alpha X^i_\omega \partial_t\phi-\beta Y^j_\omega\epsilon_j^{\ ik}\partial_tA_k\right], \label{s1}
\end{eqnarray}
where anti-symmetrization of tensor indices is denoted by square brackets and $\epsilon^{ijk}$ is the Levi-Civita tensor~\cite{MTW}.  As in the case of free-space electromagnetism~\cite{jac,wei}, the energy density (\ref{rho1}) and flux (\ref{s1}) that directly emerge from Noether's theorem are not gauge invariant. They are however equivalent to gauge-invariant quantities because they fail to be gauge invariant up to terms that \emph{identically} satisfy the conservation law (\ref{cone}). Specifically, the quantities
\begin{eqnarray}
f^{i\ t}_{\ \,t}:= -\varepsilon_0\phi E^i-\int_0^\infty\rmd\omega\,\alpha\phi X^i_\omega =:-f^{t\ i}_{\ \,t},  \\[3pt]
f^{j\ i}_{\ \,t}:=-2\kappa_0\phi \nabla^{[i}A^{j]}+\int_0^\infty\rmd\omega\,\epsilon^{ijk}\beta\phi Y_{\omega k}
\end{eqnarray}
identically satisfy
\begin{equation}  \label{fid}
\partial_t\nabla_if^{i\ t}_{\ \,t}+\nabla_i(\partial_tf^{t\ i}_{\ \,t}+\nabla_jf^{j\ i}_{\ \,t})=0.
\end{equation}
Comparing (\ref{fid}) with (\ref{cone}), we see that if $\nabla_if^{i\ t}_{\ \,t}$ is added to $\rho$, and $\partial_tf^{t\ i}_{\ \,t}+\nabla_jf^{j\ i}_{\ \,t}$ is added to $s^i$, then the conservation law (\ref{cone}) will still hold. With use of the field equations (\ref{Eeq})--(\ref{Yeq}), the energy density and flux that result from these additions are gauge invariant and are given by
\begin{eqnarray}
\fl
\rho=&\frac{\kappa_0}{2}\left[\frac{1}{c^2}\mathbf{E}^2+\mathbf{B}^2\right]  \nonumber \\[3pt]
\fl
&+\int_0^\infty\rmd\omega\left[\frac{1}{2}(\partial_t\mathbf{X}_\omega)^2+\frac{1}{2}(\partial_t\mathbf{Y}_\omega)^2+\frac{1}{2}\omega^2(\mathbf{X}_\omega^2+\mathbf{Y}_\omega^2)-\beta\mathbf{Y}_\omega\cdot\mathbf{B}\right], \label{rho} \\[3pt]
\fl
\mathbf{s}=&\kappa_0 \mathbf{E}\times \mathbf{B}-\int_0^\infty\rmd\omega\,\beta \mathbf{E}\times \mathbf{Y}_\omega. \label{s}
\end{eqnarray}
It is straightforward to verify that the conservation law (\ref{cone}) holds for (\ref{rho}) and (\ref{s}) when the fields obey the dynamical equations (\ref{Eeq})--(\ref{Yeq}).

\subsection{Momentum density and stress tensor} 
The total momentum of the electromagnetic field plus matter is conserved in flat space-time. In the description of macroscopic electromagnetism, however, the microscopic degrees of freedom of the magnetodielectric medium are not included in the action; as a result, the translation symmetry that gives rise to momentum conservation is not in general present in the dynamical system considered. A conservation law for momentum will exist within the macroscopic framework only if the coupling functions (\ref{ab}) are independent of position (a homogeneous medium) so that the action is invariant under active spatial translations of the dynamical fields. It is nevertheless instructive to retain the general case of an inhomogeneous medium; the conservation law for momentum will then be seen to fail in the inhomogeneous case due to the appearance of spatial derivatives of the coupling functions (\ref{ab}).

We make an active infinitesimal spatial translation of all the dynamical fields; in the case of the vector potential this is $\mathbf{A}(\mathbf{r},t)\rightarrow\mathbf{A} (\mathbf{r}+\mathbf{w}(\mathbf{r},t),t)$. The resulting change in the action can be written
\begin{equation} \label{actvarr}
\delta S=-\int d^4x\left(p_i\,\partial_t w^i+\sigma_i^{\ j}\,\nabla_j w^i\right),
\end{equation}
where the momentum density $\mathbf{p}$ and stress tensor $\sigma_i^{\ j}$ obey, in homogeneous media, the conservation law
\begin{equation} \label{conm}
\partial_t p_i+ \nabla_j\sigma_i^{\ j}=0.
\end{equation}
The form (\ref{actvarr}) is achieved with 
\begin{eqnarray}
\fl
p_i=&\varepsilon_0E^j\nabla_iA_j -\int_0^\infty\rmd\omega\left(\partial_t X^j_\omega\nabla_i X_{\omega j}+\partial_t Y^j_\omega\nabla_i Y_{\omega j}-\alpha X^j_\omega\nabla_iA_j\right), \label{p1} \\[3pt]
\fl
\sigma_i^{\ j}=&\mathcal{L}\,\delta_i^{\ j}+\kappa_0\left(\frac{1}{c^2}E^j\nabla_i\phi+2\nabla^{[j}A^{k]}\nabla_iA_k\right)+\int_0^\infty\rmd\omega\left(\alpha X^j_\omega\nabla_i\phi-\beta Y^k_\omega\epsilon_{k\ l}^{\ \,j}\nabla_iA^l\right), \label{stress1}
\end{eqnarray}
where $\mathcal{L}$ is the Lagrangian density, i.e.\ the integrand in the action (\ref{S}). Again, the initial results (\ref{p1}) and (\ref{stress1}) are not gauge invariant. The quantities
\begin{eqnarray}
f^{j\ t}_{\ \,i}:= -\varepsilon_0A_iE^j-\int_0^\infty\rmd\omega\,\alpha A_iX^j_\omega =:-f^{t\ j}_{\ \,i},  \\
f^{k\ j}_{\ \,i}:=-2\kappa_0A_i\nabla^{[j}A^{k]}+\int_0^\infty\rmd\omega\,\beta\epsilon^{ljk}A_iY_{\omega l},
\end{eqnarray}
identically satisfy
\begin{equation}  \label{fid2}
\partial_t\nabla_jf^{j\ t}_{\ \,i}+\nabla_j(\partial_tf^{t\ j}_{\ \,i}+\nabla_kf^{k\ j}_{\ \,i})=0.
\end{equation}
Thus, addition of $\nabla_jf^{j\ t}_{\ \,i}$ to $p_i$, and of $\partial_tf^{t\ j}_{\ \,i}+\nabla_kf^{k\ j}_{\ \,i}$ to $\sigma_i^{\ j}$, does not affect the momentum conservation law (\ref{conm}). After these additions, and use of the field equations (\ref{Eeq})--(\ref{Yeq}), we obtain a gauge-invariant momentum density and stress tensor:
\begin{eqnarray}
\fl
p_i=&\varepsilon_0(\mathbf{E}\times\mathbf{B})_i -\int_0^\infty\rmd\omega\left[\partial_t X^j_\omega\nabla_i X_{\omega j}+\partial_t Y^j_\omega\nabla_i Y_{\omega j}+\alpha (\mathbf{B}\times\mathbf{X}_\omega)_i\right], \label{p} \\[3pt]
\fl
\sigma_i^{\ j}=&\frac{1}{2}\delta_i^{\ j}(\varepsilon_0\mathbf{E}^2+\kappa_0\mathbf{B}^2)-\varepsilon_0E_iE^j-\kappa_0B_iB^j  \nonumber  \\[3pt]
\fl
&+\int_0^\infty\rmd\omega\left\{ \delta_i^{\ j}\left[\frac{1}{2}(\partial_t\mathbf{X}_\omega)^2+\frac{1}{2}(\partial_t\mathbf{Y}_\omega)^2-\frac{1}{2}\omega^2(\mathbf{X}_\omega^2+\mathbf{Y}_\omega^2)+\alpha \mathbf{X}_\omega\cdot \mathbf{E}\right]\right.  \nonumber  \\[3pt]
\fl
& \qquad \qquad \quad -\alpha E_iX^j_\omega+\beta Y_{\omega i}B^j {\Bigg\}}. \label{stress}
\end{eqnarray}
When the field equations (\ref{Eeq})--(\ref{Yeq}) hold, the momentum density (\ref{p}) and stress tensor (\ref{stress}) satisfy
\begin{equation} \label{conminhom}
\partial_t p_i+ \nabla_j\sigma_i^{\ j}=\int_0^\infty\rmd\omega\left(\mathbf{E}\cdot\mathbf{X}_\omega\nabla_i\alpha+\mathbf{B}\cdot\mathbf{Y}_\omega\nabla_i\beta\right),
\end{equation}
so that the conservation law (\ref{conm}) indeed holds for homogeneous media.

\section{Thermal equilibrium} \label{sec:thermal}
Casimir forces can be calculated from either the electromagnetic energy density or stress tensor in the presence of macroscopic media. General electromagnetic fields will exert forces on the media, but the Casimir effect is the special case when the fields are in their ground state (zero-point fields) or in thermal equilibrium with the media (the latter case of course includes the contribution of the former). To derive the Casimir effect we therefore need only assume that the eigenmodes of macroscopic QED are in a thermal mixed quantum state. The expressions for the electromagnetic energy density and stress tensor that determine the forces then follow from the general results (\ref{rho}) and (\ref{stress}) (which, as noted above, also hold for quantum field operators). This is in sharp contrast to Lifshitz theory~\cite{lif55, dzy61, LL}, where extraordinarily complicated thermodynamical and mechanical arguments are required to obtain the electromagnetic stress tensor in media and vacuum, in a manner that has nothing obvious to do with the principles of quantum mechanics~\cite{ros10}. Although the calculations below are certainly tedious, the only ingredients are the quantum theory of macroscopic electromagnetism~\cite{phi10} and a restriction to the case of thermal equilibrium.

To impose thermal equilibrium on the bosonic eigenmodes of macroscopic QED, we assume that the expectation value of the number operator of the eigenmodes is given by
\begin{eqnarray}  
\fl
\left\langle\mathbf{\hat{C}}^\dagger_\lambda(\mathbf{r},\omega)\otimes\mathbf{\hat{C}}_{\lambda'} (\mathbf{r'},\omega')\right\rangle=\mathcal{N}(\omega)\mathds{1}\delta_{\lambda \lambda'}\delta(\omega-\omega') \delta(\mathbf{r}-\mathbf{r'})  \label{CdC} \\
\qquad  \qquad \ \, =\left\langle\mathbf{\hat{C}}_{\lambda'}(\mathbf{r'},\omega')\otimes\mathbf{\hat{C}}^\dagger_{\lambda} (\mathbf{r},\omega)\right\rangle-\mathds{1}\delta_{\lambda \lambda'}\delta(\omega-\omega') \delta(\mathbf{r}-\mathbf{r'}),    \label{CCd}    \\
\quad \ \ \,     \mathcal{N}(\omega):=\left[\exp\left(\frac{\hbar\omega}{k_B T}\right)-1\right]^{-1},  \label{N} \\
\fl
\left\langle\mathbf{\hat{C}}_\lambda(\mathbf{r},\omega)\otimes\mathbf{\hat{C}}_{\lambda'} (\mathbf{r'},\omega')\right\rangle=0. \label{CC}
\end{eqnarray}
Each eigenmode at each frequency $\omega$ and each position $\mathbf{r}$ is a quantum harmonic oscillator and so the expectation value for the number of quanta (excitation level) in each of these oscillators should be given by the Planck distribution (\ref{N}). The complication in implementing this simple prescription is that the eigenmodes are a continuum in frequency and position. The obvious way of handling this last fact is to use delta functions as in (\ref{CdC}), a procedure also followed in~\cite{ros10}, where a simple quantized microscopic model of a medium was analysed. But it must be admitted that there is no clear mathematical basis for the density operator which is supposed to underlie the expectation value (\ref{CdC}). The problem is that the density operator should be defined in terms of number states of the quanta, the excitation levels of the oscillators, but because a continuum of oscillators is excited, there is no clear way of writing and normalizing these number states from which one would then construct the density operator of a thermal mixed state. This may well be an inessential technical issue, but one should bear in mind that the apparently clear intuition employed in writing (\ref{CdC})--(\ref{CC}) conceals formidable mathematical difficulties. Moreover, careful study of the derivations that follow will show that the delta function in frequency in the correlation functions (\ref{CdC}) and (\ref{CCd}) is to be regarded as a limit to be taken only in the final stages of the calculations; this allows any ambiguities arising from products of delta functions to be negotiated.

\section{Thermal field correlation functions}  \label{sec:cor}
As all  the dynamical field operators of macroscopic QED are expressible in terms of the eigenmode creation and annihilation operators, (\ref{CdC})--(\ref{CC}) immediately give the expectation values of products of field operators (correlation functions) in thermal equilibrium. For the current density operator (\ref{jopdef}) in the frequency domain we obtain
\begin{eqnarray}
\fl
\left\langle\mathbf{\hat{j}}^\dagger (\mathbf{r}, \omega)\otimes\mathbf{\hat{j}}(\mathbf{r'}, \omega')\right \rangle &=&4\pi\hbar\,\mathcal{N}(\omega)\delta(\omega-\omega')\Bigg\{\omega^2 \varepsilon_0\varepsilon_\mathrm{I}(\mathbf{r},\omega)\mathds{1}\delta(\mathbf{r}-\mathbf{r'})  \nonumber \\
&& + \left.\kappa_0\nabla\times\left[\sqrt{-\kappa_\mathrm{I}(\mathbf{r},\omega)}\,\mathds{1}\delta(\mathbf{r}-\mathbf{r'})\sqrt{-\kappa_\mathrm{I}(\mathbf{r'},\omega')}\right]\times\stackrel{\leftarrow}{\nabla'}\right\} \label{jdj} \\
\fl
&=&\frac{\mathcal{N}(\omega)}{\mathcal{N}(\omega)+1}\left\langle\mathbf{\hat{j}} (\mathbf{r'}, \omega')\otimes\mathbf{\hat{j}}^\dagger(\mathbf{r}, \omega)\right \rangle   \label{jjd} \\
\fl
\left \langle\mathbf{\hat{j}} (\mathbf{r}, \omega)\otimes\mathbf{\hat{j}}(\mathbf{r'}, \omega')\right \rangle &=&0,  \label{jj}
\end{eqnarray}
where the notation $\times\stackrel{\leftarrow}{\nabla'}$ denotes a curl with respect to the right-hand index, so that $\mathbf{V}(\mathbf{r})\times\stackrel{\leftarrow}{\nabla}= \nabla \times\mathbf{V}(\mathbf{r})$ for a vector $\mathbf{V}(\mathbf{r})$ (note that there is no minus sign included in this definition of $\nabla$ acting on the right-hand side).. 
Equations (\ref{jdj})--(\ref{jj}) can be viewed as an example of the fluctuation-dissipation theorem, but here they are a simple consequence of macroscopic QED in thermal equilibrium. 

From (\ref{Efreq}), the equal-time correlation function for the electric field operator is expressed in terms of the frequency-domain correlation function by
\begin{eqnarray}
\fl
\left\langle\mathbf{\hat{E}} (\mathbf{r}, t)\otimes\mathbf{\hat{E}}(\mathbf{r'}, t)\right \rangle=&\frac{1}{4\pi^2}\int_0^\infty\rmd\omega\int_0^\infty\rmd\omega'  \left\{ \exp[-\rmi(\omega-\omega')t]\left\langle\mathbf{\hat{E}} (\mathbf{r}, \omega)\otimes\mathbf{\hat{E}}^\dagger(\mathbf{r'}, \omega')\right \rangle\right.  \nonumber \\
\fl
&\left.+\exp[\rmi(\omega-\omega')t]\left\langle\mathbf{\hat{E}}^ \dagger (\mathbf{r}, \omega)\otimes\mathbf{\hat{E}}(\mathbf{r'}, \omega')\right \rangle\right\}.  \label{EEEEf}
\end{eqnarray}
(Terms of the form $\langle\mathbf{\hat{E}}(\omega)\mathbf{\hat{E}}(\omega') \rangle$ and $\langle\mathbf{\hat{E}}^\dagger(\omega)\mathbf{\hat{E}}^\dagger(\omega') \rangle$ vanish.) The frequency-domain correlation functions are written in terms of the Green bi-tensor using (\ref{Efreq}), (\ref{EopG}) and (\ref{jdj})--(\ref{jj}); after spatial integrations by parts we obtain
\begin{eqnarray}
\fl
\left\langle\hat{E}^\dagger_i (\mathbf{r}, \omega)\hat{E}_j(\mathbf{r'}, \omega')\right \rangle          \nonumber \\  
 = 4\pi\hbar\mu_0 \int\rmd^3\mathbf{r''} \,\mathcal{N}(\omega)\omega^2\delta(\omega-\omega')\left\{\frac{\omega^2}{c^2}   \varepsilon_\mathrm{I}(\mathbf{r''},\omega)G_{ik}(\mathbf{r},\mathbf{r''},\omega)G^{*\ k}_{\ j}(\mathbf{r'},\mathbf{r''},\omega)  \right.      \nonumber \\
\quad  \left.  -\kappa_\mathrm{I}(\mathbf{r''},\omega)[\mathbf{G}(\mathbf{r},\mathbf{r''},\omega)\times\stackrel{\leftarrow}{\nabla''}]_{ik}\,[\mathbf{G}^*(\mathbf{r'},\mathbf{r''},\omega)\times\stackrel{\leftarrow}{\nabla''}]_{j}^{\ k}\right\}.     \label{EdE1} 
\end{eqnarray}
The correlation function $\langle\hat{E}_i (\mathbf{r}, \omega)\hat{E}^\dagger_j(\mathbf{r'}, \omega')\rangle$ is given by the complex conjugate of (\ref{EdE1}) with $\mathcal{N}(\omega)$ replaced by $\mathcal{N}(\omega)+1$. The right-hand side of (\ref{EdE1}) can be simplified as follows. We first note the symmetry property of the Green bi-tensor that holds for media invariant under time reversal (non-magnetic,\footnote{Note the distinction between the somewhat confusing terms {\it magnetic media} and {\it magnetodielectric media}. The former refers to media with permanent magnetizations in the absence of applied fields, the latter to media that have a non-trivial magnetic (and electric) response to electromagnetic fields.} non-moving media)~\cite{LL}:
\begin{equation}   \label{recip}
G_{ij}(\mathbf{r},\mathbf{r'},\omega)=G_{ji}(\mathbf{r'},\mathbf{r},\omega).
\end{equation}
Take the matrix product of $\mathbf{G}^*(\mathbf{r''},\mathbf{r},\omega)$ with (\ref{green}) and integrate over $\mathbf{r}$; after integration by parts and use of (\ref{recip}), the imaginary part of the resulting relation simplifies to
\begin{eqnarray}
\fl
\int\!
\rmd^3\mathbf{r}\left\{-\kappa_\mathrm{I}(\mathbf{r},\omega)[\mathbf{G}^*(\mathbf{r''},\mathbf{r},\omega)\times\stackrel{\leftarrow}{\nabla}]_{ik}\,[\nabla \times\mathbf{G}(\mathbf{r},\mathbf{r'},\omega)]_{\ j}^{k}  \right.    \nonumber \\
\left. +\frac{\omega^2}{c^2}   \varepsilon_\mathrm{I}(\mathbf{r},\omega)G^*_{\ ik}(\mathbf{r''},\mathbf{r},\omega)G^{k}_{\ j}(\mathbf{r},\mathbf{r'},\omega)\right\}=G_{\mathrm{I} ij}(\mathbf{r''},\mathbf{r'},\omega).   \label{Gid}
\end{eqnarray}
Use of (\ref{Gid}) and its complex conjugate, together with (\ref{recip}), allows (\ref{EdE1}) to be simplified to
\begin{eqnarray}
\left\langle\mathbf{\hat{E}}^ \dagger (\mathbf{r}, \omega)\otimes\mathbf{\hat{E}}(\mathbf{r'}, \omega')\right \rangle \!&=& 4\pi\hbar\mu_0 \delta(\omega-\omega')\mathcal{N}(\omega)\omega^2\mathbf{G}_{\mathrm{I}}(\mathbf{r},\mathbf{r'},\omega)        \label{EdE}  \\
&=&\frac{\mathcal{N}(\omega)}{\mathcal{N}(\omega)+1}\left\langle\mathbf{\hat{E}} (\mathbf{r}, \omega)\otimes\mathbf{\hat{E}}^\dagger(\mathbf{r'}, \omega')\right \rangle.  \label{EEd}
\end{eqnarray}
The equal-time correlation function (\ref{EEEEf}) is, from (\ref{EdE}) and (\ref{EEd}),
\begin{equation}  \label{EE}
\left\langle\mathbf{\hat{E}} (\mathbf{r}, t)\otimes\mathbf{\hat{E}}(\mathbf{r'}, t)\right \rangle=\frac{\hbar\mu_0}{\pi}\int_0^\infty\rmd\omega\,\omega^2\coth\left(\frac{\hbar\omega}{2k_B T}\right)\mathbf{G}_{\mathrm{I}}(\mathbf{r},\mathbf{r'},\omega),
\end{equation}
where a factor of $2\mathcal{N}(\omega)+1$ has been rewritten as a hyperbolic cotangent (recall (\ref{N})). Correlation functions for the magnetic field operator are easily found from the electric-field expressions through use of (\ref{BE}), with the results
\begin{eqnarray}
\fl
\left\langle\mathbf{\hat{B}}^ \dagger (\mathbf{r}, \omega)\otimes\mathbf{\hat{B}}(\mathbf{r'}, \omega')\right \rangle \!&=& 4\pi\hbar\mu_0 \delta(\omega-\omega')\mathcal{N}(\omega)\nabla\times\mathbf{G}_{\mathrm{I}}(\mathbf{r},\mathbf{r'},\omega)\times\stackrel{\leftarrow}{\nabla'}        \label{BdB}  \\
&=&\frac{\mathcal{N}(\omega)}{\mathcal{N}(\omega)+1}\left\langle\mathbf{\hat{B}} (\mathbf{r}, \omega)\otimes\mathbf{\hat{B}}^\dagger(\mathbf{r'}, \omega')\right \rangle,  \label{BBd}   \\
\fl
\left\langle\mathbf{\hat{B}} (\mathbf{r}, t)\otimes\mathbf{\hat{B}}(\mathbf{r'}, t)\right \rangle&=&\frac{\hbar\mu_0}{\pi}\int_0^\infty\rmd\omega \,\coth\left(\frac{\hbar\omega}{2k_B T}\right)\nabla\times\mathbf{G}_{\mathrm{I}}(\mathbf{r},\mathbf{r'},\omega)\times\stackrel{\leftarrow}{\nabla'}.  \label{BB}  
\end{eqnarray}

Correlation functions for the reservoir field operators $\mathbf{\hat{X}}_\omega$ and $\mathbf{\hat{Y}}_\omega$ are found using (\ref{XCE}) and (\ref{YCB}). Our goal is to calculate the electromagnetic part of the energy density and stress tensor, by isolating the electromagnetic part of the expressions (\ref{rho}) and (\ref{stress}) when they are evaluated in the case of thermal equilibrium. This requires us to eliminate the reservoir fields by expressing them in terms of the electromagnetic fields and free-field terms that are independent of the coupling to the electromagnetic fields. Equations (\ref{XCE}) and (\ref{YCB}) perform this separation of the reservoir fields into terms that would be present in the absence of any coupling to the electromagnetic fields (the first terms on the right-hand sides) and electromagnetic terms (the second terms on the right-hand sides). Only terms in correlation functions that depend on the electromagnetic fields are required to obtain the electromagnetic part of the energy density and stress tensor. We therefore drop terms in correlation functions that are independent of the electromagnetic fields. (Further remarks on the non-electromagnetic parts of the energy density and stress tensor will be made in the next section.)

The correlation function in the frequency domain of the reservoir field $\mathbf{\hat{X}}_\omega$ with itself is obtained using (\ref{XCE}). We use a subscript $\scriptstyle E$ on the correlation function to denote the fact that we include only the terms that depend on the electric field; this yields the following correlation function:
\begin{eqnarray}
\fl
\left\langle\mathbf{\hat{X}}^\dagger_\omega (\mathbf{r}, \omega')\otimes\mathbf{\hat{X}}_\omega(\mathbf{r'}, \omega'')\right \rangle_E     \nonumber \\[3pt]
\fl
\quad =\sqrt{\frac{\hbar}{2\omega}}\delta(\omega-\omega')\frac{\pi\alpha(\mathbf{r'},\omega)}{\omega}\left[\frac{1}{\omega-\omega''-\rmi0^+}+\frac{1}{\omega+\omega''}\right]\left\langle\mathbf{\hat{C}}^\dagger_\mathrm{e} (\mathbf{r}, \omega')\otimes\mathbf{\hat{E}}(\mathbf{r'}, \omega'')\right\rangle      \nonumber \\[3pt]
\fl
\qquad +\sqrt{\frac{\hbar}{2\omega}}\delta(\omega-\omega'')\frac{\pi\alpha(\mathbf{r'},\omega)}{\omega}\left[\frac{1}{\omega-\omega'+\rmi0^+}+\frac{1}{\omega+\omega'}\right]\left\langle\mathbf{\hat{E}}^\dagger(\mathbf{r}, \omega')\otimes\mathbf{\hat{C}}_\mathrm{e}(\mathbf{r'}, \omega'')\right\rangle      \nonumber \\[3pt]
\fl
\qquad +\frac{\alpha(\mathbf{r},\omega)\alpha(\mathbf{r'},\omega)}{4\omega^2}\left[\frac{1}{\omega-\omega'+\rmi0^+}+\frac{1}{\omega+\omega'}\right]\left[\frac{1}{\omega-\omega''-\rmi0^+}+\frac{1}{\omega+\omega''}\right] \nonumber \\[3pt]
\fl
\qquad\quad \times\left\langle\mathbf{\hat{E}}^\dagger(\mathbf{r}, \omega')\otimes\mathbf{\hat{E}}(\mathbf{r'}, \omega'')\right\rangle.
\label{XdX0}
\end{eqnarray}
The third term on the right-hand side of (\ref{XdX0}) will produce a product of principal values if the quantities containing the infinitesimal number $0^+$ are expanded in terms of principal values and delta functions. This product of principal values would need to be treated with care depending on the integration variable to which they are referred (as well as integrations over $\omega'$ and $\omega''$, there will be an integration over $\omega$ in the energy density and stress). It is safer to rewrite the term in question in a form that does not produce a product of principal values; this can be done using the identity 
\begin{eqnarray}
\fl
\frac{1}{(\omega-\omega'+\rmi0^+)(\omega-\omega''-\rmi0^+)}=\frac{1}{\omega'-\omega''-2\rmi0^+}\left(\frac{1}{\omega-\omega'+\rmi0^+}-\frac{1}{\omega-\omega''-\rmi0^+}\right)   \nonumber \\[3pt] 
=\frac{1}{\omega'-\omega''-2\rmi0^+}\left[\mathrm{P}\frac{\omega'-\omega''}{(\omega-\omega')(\omega-\omega'')}-\rmi\pi\delta(\omega-\omega')-\rmi\pi\delta(\omega-\omega'')\right]   \nonumber \\[3pt] 
=\mathrm{P}\frac{1}{(\omega-\omega')(\omega-\omega'')}-\rmi\pi\frac{\delta(\omega-\omega')+\delta(\omega-\omega'')}{\omega'-\omega''-2\rmi0^+}.    \label{PVid}
\end{eqnarray}
From (\ref{PVid}), the product of the quantities in square brackets in the third term on the right-hand side of (\ref{XdX0}) simplifies to
\begin{eqnarray}
\fl
\left[\frac{1}{\omega-\omega'+\rmi0^+}+\frac{1}{\omega+\omega'}\right]\left[\frac{1}{\omega-\omega''-\rmi0^+}+\frac{1}{\omega+\omega''}\right]   \nonumber \\[3pt] 
=\mathrm{P}\frac{4\omega^2}{(\omega^2-\omega'^2)(\omega^2-\omega''^2)}-\rmi\pi\frac{\delta(\omega-\omega')+\delta(\omega-\omega'')}{\omega'-\omega''-2\rmi0^+},  \label{PVid2}
\end{eqnarray}
where all terms apart from the right-hand side of (\ref{PVid}) were expanded as principal values and delta functions (we do not perform a similar expansion in the last term in (\ref{PVid}) and (\ref{PVid2}) at this stage simply to save space). At the risk of long-windedness, we rewrite (\ref{XdX0}) having inserted (\ref{PVid2}):
\begin{eqnarray}
\fl
\left\langle\mathbf{\hat{X}}^\dagger_\omega (\mathbf{r}, \omega')\otimes\mathbf{\hat{X}}_\omega(\mathbf{r'}, \omega'')\right \rangle_E     \nonumber \\[3pt]
\fl
\quad =\sqrt{\frac{\hbar}{2\omega}}\delta(\omega-\omega')\frac{\pi\alpha(\mathbf{r'},\omega)}{\omega}\left[\frac{1}{\omega-\omega''-\rmi0^+}+\frac{1}{\omega+\omega''}\right]\left\langle\mathbf{\hat{C}}^\dagger_\mathrm{e} (\mathbf{r}, \omega')\otimes\mathbf{\hat{E}}(\mathbf{r'}, \omega'')\right\rangle      \nonumber \\[3pt]
\fl
\qquad +\sqrt{\frac{\hbar}{2\omega}}\delta(\omega-\omega'')\frac{\pi\alpha(\mathbf{r'},\omega)}{\omega}\left[\frac{1}{\omega-\omega'+\rmi0^+}+\frac{1}{\omega+\omega'}\right]\left\langle\mathbf{\hat{E}}^\dagger(\mathbf{r}, \omega')\otimes\mathbf{\hat{C}}_\mathrm{e}(\mathbf{r'}, \omega'')\right\rangle      \nonumber \\[3pt]
\fl
\qquad +\frac{\alpha(\mathbf{r},\omega)\alpha(\mathbf{r'},\omega)}{4\omega^2}\left[\mathrm{P}\frac{4\omega^2}{(\omega^2-\omega'^2)(\omega^2-\omega''^2)}-\rmi\pi\frac{\delta(\omega-\omega')+\delta(\omega-\omega'')}{\omega'-\omega''-2\rmi0^+}\right] \nonumber \\[3pt]
\fl
\qquad\quad \times\left\langle\mathbf{\hat{E}}^\dagger(\mathbf{r}, \omega')\otimes\mathbf{\hat{E}}(\mathbf{r'}, \omega'')\right\rangle.
\label{XdX0.1}
\end{eqnarray}
The correlation function in the third term on the right-hand side of (\ref{XdX0.1}) is given by (\ref{EdE}); the correlation functions in the first two terms are shown by (\ref{CdC})--(\ref{CC}), (\ref{EopG}) and (\ref{jopdef}) to be
\begin{eqnarray}
\fl
\left\langle\hat{C}^\dagger_{\mathrm{e}i} (\mathbf{r}, \omega')\hat{E}_j(\mathbf{r'}, \omega'')\right\rangle=2\pi\mu_0\omega''^2\mathcal{N}(\omega')\left[\frac{\hbar\varepsilon_0}{\pi}\varepsilon_\mathrm{I}(\mathbf{r},\omega'')\right]^{1/2}G_{ji}(\mathbf{r'},\mathbf{r},\omega'')\delta(\omega'-\omega''),  \label{CdE} \\[3pt]
\fl
\left\langle\hat{E}^\dagger_{i} (\mathbf{r}, \omega')\hat{C}_{\mathrm{e}j}(\mathbf{r'}, \omega'')\right\rangle=2\pi\mu_0\omega'^2\mathcal{N}(\omega')\left[\frac{\hbar\varepsilon_0}{\pi}\varepsilon_\mathrm{I}(\mathbf{r'},\omega')\right]^{1/2}G^*_{ij}(\mathbf{r},\mathbf{r'},\omega')\delta(\omega'-\omega'').  \label{EdC} 
\end{eqnarray}

Now we must focus on the expectation values containing only $\mathbf{\hat{X}}_\omega$ that are required for the expectation values of the energy density (\ref{rho}) and stress tensor (\ref{stress}). In writing the expectation values we employ a shortened notation for two limits involving the Green bi-tensor:
\begin{eqnarray}
\Delta^{\!E}_{\ \,ij}(\mathbf{r},\omega):=\omega^2\lim_{\mathbf{r'}\to\mathbf{r}}G_{ij}(\mathbf{r},\mathbf{r'},\omega),   \label{DeltaE}  \\
\Delta^{\!B}_{\ \,ij}(\mathbf{r},\omega):=\lim_{\mathbf{r'}\to\mathbf{r}}[\nabla\times\mathbf{G}(\mathbf{r},\mathbf{r'},\omega)\times\stackrel{\leftarrow}{\nabla'}]_{ij}.   \label{DeltaB}
\end{eqnarray}
The limit $\mathbf{r'}\to\mathbf{r}$ appears because the expectation values of the energy density (\ref{rho}) and stress tensor (\ref{stress}) contain correlation functions evaluated at $\mathbf{r'}=\mathbf{r}$. But the Green bi-tensor itself diverges when $\mathbf{r'}=\mathbf{r}$, so the limit in (\ref{DeltaE}) and (\ref{DeltaB}) must be understood to be taken only in the final expressions for physical quantities. The zero-point part of the Casimir effect requires a regularization to remove the divergent zero-point energy that is always present in a homogeneous medium (including vacuum) and that does not contribute to the Casimir force. This regularization is implemented at the level of the Green bi-tensor in a manner familiar from Lifshitz theory~\cite{lif55, dzy61, LL}. We also understand this regularization to be included in (\ref{DeltaE}) and (\ref{DeltaB}) when they appear below in the Casimir energy density and stress tensor.

For the energy density (\ref{rho}) we need $\int_0^\infty\rmd\omega\langle(\partial\mathbf{\hat{X}}_\omega)^2/2+\omega^2\mathbf{\hat{X}}_\omega^2/2\rangle$; from the general relation (\ref{EEEEf}) it therefore follows that 
we must insert a factor of $(\omega'\omega''+\omega^2)/2$ into (\ref{XdX0.1}). Inserting this factor and substituting (\ref{CdE}), (\ref{EdC}), (\ref{EdE}) and (\ref{ab}) we obtain, after minor simplifications,
\begin{eqnarray}
\fl
\left\langle\frac{1}{2}(\omega'\omega''+\omega^2)\hat{X}^\dagger_{\omega i} (\mathbf{r}, \omega'){\hat{X}}_{\omega j}(\mathbf{r}, \omega'')\right \rangle_E     \nonumber \\[3pt]
\fl
\quad =\frac{\hbar\pi}{c^2}\frac{(\omega'\omega''+\omega^2)}{\omega}\sqrt{ \varepsilon_\mathrm{I}(\mathbf{r},\omega) \varepsilon_\mathrm{I}(\mathbf{r},\omega'')}\left[\mathrm{P}\frac{2\omega}{\omega^2-\omega''^2}+\rmi\pi\delta(\omega-\omega'')\right]\mathcal{N}(\omega')   \nonumber \\[3pt]
\times\delta(\omega-\omega')\delta(\omega'-\omega'')\Delta^{\!E}_{\ \,ji}(\mathbf{r},\omega')      \nonumber \\[3pt]
\fl
\qquad +\frac{\hbar\pi}{c^2}\frac{(\omega'\omega''+\omega^2)}{\omega}\sqrt{ \varepsilon_\mathrm{I}(\mathbf{r},\omega) \varepsilon_\mathrm{I}(\mathbf{r},\omega')}\left[\mathrm{P}\frac{2\omega}{\omega^2-\omega'^2}-\rmi\pi\delta(\omega-\omega')\right]\mathcal{N}(\omega')   \nonumber \\[3pt]
\times\delta(\omega-\omega'')\delta(\omega'-\omega'')\Delta^{\!E*}_{\ \,ij}(\mathbf{r},\omega')      \nonumber \\[3pt]
\fl
\qquad  +\frac{\hbar}{c^2}\frac{(\omega'\omega''+\omega^2)}{\omega}\varepsilon_\mathrm{I}(\mathbf{r},\omega)\left[\mathrm{P}\frac{4\omega^2}{(\omega^2-\omega'^2)(\omega^2-\omega''^2)}-\rmi\pi\frac{\delta(\omega-\omega')+\delta(\omega-\omega'')}{\omega'-\omega''-2\rmi0^+}\right] \nonumber \\[3pt]
\times\mathcal{N}(\omega')\delta(\omega'-\omega'')\mathrm{Im}\Delta^{\!E}_{\ \,ij}(\mathbf{r},\omega') .
\label{XdX0.2}
\end{eqnarray}
The right-hand side of (\ref{XdX0.2}) consists of a sum of three terms. The first two terms each contain a product of three delta functions; these terms containing three delta functions can be combined and written as
\begin{eqnarray}
\fl
\frac{2\hbar\pi^2\omega}{c^2}\varepsilon_\mathrm{I}(\mathbf{r},\omega)\mathcal{N}(\omega)\left[\rmi\Delta^{\!E}_{\ \,ji}(\mathbf{r},\omega')-\rmi \Delta^{\!E*}_{\ \,ij}(\mathbf{r},\omega')\right] \delta(\omega-\omega')\delta(\omega-\omega'')\delta(\omega'-\omega'').  \label{3delta}
\end{eqnarray}
The third term on the right-hand side of (\ref{XdX0.2}) has a part containing the sum of two delta functions in a numerator; this part gives the contribution
\begin{eqnarray}
\fl
-\rmi\frac{\hbar\pi}{c^2}\frac{(\omega'\omega''+\omega^2)}{\omega}\varepsilon_\mathrm{I}(\mathbf{r},\omega)[\delta(\omega-\omega')+\delta(\omega-\omega'')]\left[\mathrm{P}\frac{1}{(\omega'-\omega'')}+\rmi\pi\delta(\omega'-\omega'')\right]
\nonumber \\[3pt]
\times\mathcal{N}(\omega')\delta(\omega'-\omega'')\mathrm{Im}\Delta^{\!E}_{\ \,ij}(\mathbf{r},\omega')   \nonumber \\[3pt]
\fl
\qquad =-\rmi\frac{\hbar\pi}{c^2}\frac{(\omega'\omega''+\omega^2)}{\omega}\varepsilon_\mathrm{I}(\mathbf{r},\omega)\left[\mathrm{P}\frac{\delta(\omega-\omega')}{(\omega-\omega'')}+\mathrm{P}\frac{\delta(\omega-\omega'')}{(\omega'-\omega)}+2\rmi\pi\delta(\omega-\omega')\delta(\omega-\omega'')\right]     \nonumber \\[3pt]
\qquad  \times\mathcal{N}(\omega')\delta(\omega'-\omega'')\mathrm{Im}\Delta^{\!E}_{\ \,ij}(\mathbf{r},\omega')    \nonumber \\[3pt]
\fl
\qquad =\frac{4\hbar\pi^2\omega}{c^2}\varepsilon_\mathrm{I}(\mathbf{r},\omega)\mathcal{N}(\omega)\mathrm{Im}\Delta^{\!E}_{\ \,ij}(\mathbf{r},\omega') \delta(\omega-\omega')\delta(\omega-\omega'')\delta(\omega'-\omega''),
\end{eqnarray}
which cancels with (\ref{3delta}). The first two terms in the sum on the right-hand of (\ref{XdX0.2}) have now been reduced to the parts containing a principal value; consider the first of these, namely
\begin{equation}
\fl
\frac{2\hbar\pi}{c^2}(\omega'\omega''+\omega^2)\sqrt{ \varepsilon_\mathrm{I}(\mathbf{r},\omega) \varepsilon_\mathrm{I}(\mathbf{r},\omega'')}\,\mathrm{P}\frac{1}{\omega^2-\omega''^2}\mathcal{N}(\omega')\delta(\omega-\omega')\delta(\omega'-\omega'')\Delta^{\!E}_{\ \,ji}(\mathbf{r},\omega').  \label{P2del}
\end{equation}
Because of the delta functions, (\ref{P2del}) is restricted to contribute only when the denominator in the principal value vanishes; by the definition of a principal value there is no such contribution, unless the factor $\omega-\omega''$ in the denominator $\omega^2-\omega''^2$ is canceled by an equal factor $\omega-\omega''$ in the numerator. But there is such a factor in the numerator, obtained by Taylor expanding the remaining function of $\omega$ (apart from the the factor $\omega-\omega''$ in the denominator) around the point $\omega=\omega''$; the contribution of (\ref{P2del}) is therefore
\begin{eqnarray}
\fl
\frac{2\hbar\pi}{c^2}\sqrt{ \varepsilon_\mathrm{I}(\mathbf{r},\omega'') }\mathcal{N}(\omega')\delta(\omega-\omega')\delta(\omega'-\omega'')\Delta^{\!E}_{\ \,ji}(\mathbf{r},\omega')\frac{\rmd}{\rmd\omega}\left.\left[\sqrt{ \varepsilon_\mathrm{I}(\mathbf{r},\omega)}\frac{(\omega'\omega''+\omega^2)}{\omega+\omega''}\right]\right|_{\omega=\omega''}  \nonumber \\[3pt]
= \frac{\hbar\pi}{c^2}\frac{\rmd\left[\omega''\varepsilon_\mathrm{I}(\mathbf{r},\omega'')\right]}{\rmd\omega''} \mathcal{N}(\omega')\delta(\omega-\omega')\delta(\omega'-\omega'')\Delta^{\!E}_{\ \,ji}(\mathbf{r},\omega').             \label{P2del3}
\end{eqnarray}
The principal value in the second term in the sum on the right-hand of (\ref{XdX0.2}) is dealt with in the manner employed for (\ref{P2del}). Implementing all these simplifications of (\ref{XdX0.2}), and using the symmetry $\Delta^{\!E}_{\ \,ij}(\mathbf{r},\omega)=\Delta^{\!E}_{\ \,ji}(\mathbf{r},\omega)$ that follows from (\ref{recip}) and (\ref{DeltaE}), we obtain
\begin{eqnarray}
\fl
\left\langle\frac{1}{2}(\omega'\omega''+\omega^2)\hat{X}^\dagger_{\omega i} (\mathbf{r}, \omega'){\hat{X}}_{\omega j}(\mathbf{r}, \omega'')\right \rangle_E     \nonumber \\[3pt]
=\frac{2\hbar\pi}{c^2}\frac{\rmd\left[\omega'\varepsilon_\mathrm{I}(\mathbf{r},\omega')\right]}{\rmd\omega'} \mathcal{N}(\omega')\delta(\omega-\omega')\delta(\omega'-\omega'')\mathrm{Re}\Delta^{\!E}_{\ \,ij}(\mathbf{r},\omega')     \nonumber \\[3pt]
\quad  +\frac{4\hbar}{c^2}\mathrm{P}\frac{\omega(\omega'^2+\omega^2)}{(\omega^2-\omega'^2)^2}
\varepsilon_\mathrm{I}(\mathbf{r},\omega)\mathcal{N}(\omega')\delta(\omega'-\omega'')\mathrm{Im}\Delta^{\!E}_{\ \,ij}(\mathbf{r},\omega') .
\label{XdXen}     \\[3pt]
=\frac{\mathcal{N}(\omega')}{\mathcal{N}(\omega')+1}\left\langle\frac{1}{2}(\omega'\omega''+\omega^2)\hat{X}_{\omega i} (\mathbf{r}, \omega'){\hat{X}}^\dagger_{\omega j}(\mathbf{r}, \omega'')\right \rangle_E.  \label{XXden}
\end{eqnarray}
The equality (\ref{XXden}) is easily verified by tracing back minor changes in the derivation of (\ref{XdXen}). We can now write the time-domain expectation value $\int_0^\infty\rmd\omega\langle(\partial\mathbf{\hat{X}}_\omega)^2/2+\omega^2\mathbf{\hat{X}}_\omega^2/2\rangle$ using (\ref{XXden}), (\ref{XdXen}) and the general relation (\ref{EEEEf}):
\begin{eqnarray}
\fl
\left\langle\int_0^\infty\rmd\omega\left[\frac{1}{2}(\partial_t\mathbf{\hat{X}}_\omega)^2+\frac{1}{2}\omega^2\mathbf{\hat{X}}_\omega^2\right]\right\rangle_E   \nonumber \\[3pt]
\fl
\qquad \quad  =\frac{\hbar}{2\pi c^2}\int_0^\infty\rmd\omega\,\frac{\rmd\left[\omega\varepsilon_\mathrm{I}(\mathbf{r},\omega)\right]}{\rmd\omega}  \coth\left(\frac{\hbar\omega}{2k_B T}\right)\mathrm{Re}\Delta^{\!E\ i}_{\ \,i}(\mathbf{r},\omega) \nonumber \\[3pt]
\fl
\qquad \qquad  +\frac{\hbar}{\pi^2c^2}\int_0^\infty\rmd\omega\int_0^\infty\rmd\omega' \,\mathrm{P}\frac{\omega(\omega'^2+\omega^2)}{(\omega^2-\omega'^2)^2} \varepsilon_\mathrm{I}(\mathbf{r},\omega)  \coth\left(\frac{\hbar\omega'}{2k_B T}\right)\mathrm{Im}\Delta^{\!E\ i}_{\ \,i}(\mathbf{r},\omega')     \label{rhoX1}
\end{eqnarray}
Use of the identity
\begin{equation}
\frac{\omega(\omega'^2+\omega^2)}{(\omega^2-\omega'^2)^2}=\frac{\rmd}{\rmd\omega'}\left(\frac{\omega\omega'}{\omega^2-\omega'^2}\right)
\end{equation}
in the last term of (\ref{rhoX1}) allows the integration over $\omega$ to be performed by means of the Kramers-Kronig relation (\ref{KK}), yielding finally
\begin{eqnarray}
\fl
\left\langle\int_0^\infty\rmd\omega\left[\frac{1}{2}(\partial_t\mathbf{\hat{X}}_\omega)^2+\frac{1}{2}\omega^2\mathbf{\hat{X}}_\omega^2\right]\right\rangle _E  \nonumber \\[3pt]
\fl
\qquad \qquad  =\frac{\hbar}{2\pi c^2}\mathrm{Im}\int_0^\infty\rmd\omega' \, \coth\left(\frac{\hbar\omega'}{2k_B T}\right)\left(\frac{\rmd}{\rmd\omega'}\left\{\omega'\left[\varepsilon(\mathbf{r},\omega')-1\right]\right\}\right) \Delta^{\!E\ i}_{\ \,i}(\mathbf{r},\omega').    \label{rhoX}
\end{eqnarray}

Turning now to the the expectation value containing only $\mathbf{\hat{X}}_\omega$ that appears in the expectation value of the stress tensor (\ref{stress}), we see that this is $\int_0^\infty\rmd\omega\langle(\partial\mathbf{\hat{X}}_\omega)^2/2-\omega^2\mathbf{\hat{X}}_\omega^2/2\rangle$. Recalling the general relation (\ref{EEEEf}), we must therefore insert a factor  $(\omega'\omega''-\omega^2)/2$ in (\ref{XdX0.1}). This factor, in the first term on the right-hand side of (\ref{XdX0.1}), can be written $\omega'(\omega''-\omega)/2$ because of the presence of $\delta(\omega-\omega')$, and so the pole at $\omega=\omega''$ inside the square brackets is cancelled. Similarly, the pole at $\omega=\omega'$ in the second term  on the right-hand side of (\ref{XdX0.1}) is also cancelled by the factor $(\omega'\omega''-\omega^2)/2$. In the final term on the right-hand side of (\ref{XdX0.1}), multiplication by the factor $(\omega'\omega''-\omega^2)/2$ also produces a simplification that follows from
\begin{equation}
\fl
(\omega'\omega''-\omega^2)\frac{\delta(\omega-\omega')+\delta(\omega-\omega'')}{\omega'-\omega''-2\rmi0^+}=-\omega'\delta(\omega-\omega')+\omega''\delta(\omega-\omega''),
\end{equation}
which vanishes when combined with a factor $\delta(\omega'-\omega'')$ from (\ref{EdE}).
We substitute (\ref{CdE}), (\ref{EdC}), (\ref{EdE}) and (\ref{ab}) and find
\begin{eqnarray}
\fl
\left\langle\frac{1}{2}(\omega'\omega''-\omega^2)\hat{X}^\dagger_{\omega i} (\mathbf{r}, \omega'){\hat{X}}_{\omega j}(\mathbf{r}, \omega'')\right \rangle_E     \nonumber \\[3pt]
 =-\frac{2\hbar\pi}{c^2}\varepsilon_\mathrm{I}(\mathbf{r},\omega)\mathcal{N}(\omega') \delta(\omega-\omega')\delta(\omega'-\omega'')\mathrm{Re}\Delta^{\!E}_{\ \,ij}(\mathbf{r},\omega')      \nonumber \\[3pt]
\quad -\frac{4\hbar}{c^2}\varepsilon_\mathrm{I}(\mathbf{r},\omega)\mathrm{P}\frac{\omega}{(\omega^2-\omega'^2)}\mathcal{N}(\omega')\delta(\omega'-\omega'')\mathrm{Im}\Delta^{\!E}_{\ \,ij}(\mathbf{r},\omega') 
\label{XdXstress}     \\[3pt]
=\frac{\mathcal{N}(\omega')}{\mathcal{N}(\omega')+1}\left\langle\frac{1}{2}(\omega'\omega''-\omega^2)\hat{X}_{\omega i} (\mathbf{r}, \omega'){\hat{X}}^\dagger_{\omega j}(\mathbf{r}, \omega'')\right \rangle_E.  \label{XXdstress}
\end{eqnarray}
The time-domain expectation value $\int_0^\infty\rmd\omega\langle(\partial\mathbf{\hat{X}}_\omega)^2/2-\omega^2\mathbf{\hat{X}}_\omega^2/2\rangle$ follows from (\ref{XdXstress}), (\ref{XXdstress}) and the general relation (\ref{EEEEf}); the Kramers-Kronig relation (\ref{KK}) can be immediately applied, with the result
\begin{eqnarray}
\fl
\left\langle\int_0^\infty\rmd\omega\left[\frac{1}{2}(\partial_t\mathbf{\hat{X}}_\omega)^2-\frac{1}{2}\omega^2\mathbf{\hat{X}}_\omega^2\right]\right\rangle_E   \nonumber \\[3pt]
\fl
\qquad \qquad  =-\frac{\hbar}{2\pi c^2}\mathrm{Im}\int_0^\infty\rmd\omega' \, \coth\left(\frac{\hbar\omega'}{2k_B T}\right)\left[\varepsilon(\mathbf{r},\omega')-1\right]\Delta^{\!E\ i}_{\ \,i}(\mathbf{r},\omega').    \label{stressX}
\end{eqnarray}

Expectation values for the $\mathbf{\hat{Y}}_\omega$ field analogous to (\ref{rhoX}) and (\ref{stressX}) are also required.  In keeping with the discussion earlier in this section, only terms that depend on the magnetic-field part of $\mathbf{\hat{Y}}_\omega$ in (\ref{YCB}) are included in the calculation and we denote this fact by a subscript $\scriptstyle B$ on expectation values. The derivations are very similar to those described in detail above for the case of the $\mathbf{\hat{X}}_\omega$ field; we therefore simply state the results:
\begin{eqnarray}
\fl
\left\langle\int_0^\infty\rmd\omega\left[\frac{1}{2}(\partial_t\mathbf{\hat{Y}}_\omega)^2+\frac{1}{2}\omega^2\mathbf{\hat{Y}}_\omega^2\right]\right\rangle_B   \nonumber \\[3pt]
\fl
\qquad \qquad  =\frac{\hbar}{2\pi}\mathrm{Im}\int_0^\infty\rmd\omega' \, \coth\left(\frac{\hbar\omega'}{2k_B T}\right)\left(\frac{\rmd}{\rmd\omega'}\left\{-\omega'\left[\kappa(\mathbf{r},\omega')-1\right]\right\}\right)\Delta^{\!B\ i}_{\ \,i}(\mathbf{r},\omega').    \label{rhoY}
\end{eqnarray}
\begin{eqnarray}
\fl
\left\langle\int_0^\infty\rmd\omega\left[\frac{1}{2}(\partial_t\mathbf{\hat{Y}}_\omega)^2-\frac{1}{2}\omega^2\mathbf{\hat{Y}}_\omega^2\right]\right\rangle_B   \nonumber \\[3pt]
\fl
\qquad \qquad  =\frac{\hbar}{2\pi}\mathrm{Im}\int_0^\infty\rmd\omega' \, \coth\left(\frac{\hbar\omega'}{2k_B T}\right)\left[\kappa(\mathbf{r},\omega')-1\right]\Delta^{\!B\ i}_{\ \,i}(\mathbf{r},\omega'),    \label{stressY}
\end{eqnarray}
where the definition (\ref{DeltaB}) has been employed.

To calculate the expectation values of the energy density (\ref{rho}) and stress tenor (\ref{stress}) we also require the expectation values of $\int_0^\infty\rmd\omega\,\beta (\mathbf{r},\omega)\hat{Y}_{\omega i}\hat{B}_j$ and $\int_0^\infty\rmd\omega\,\alpha (\mathbf{r},\omega)\hat{X}_{\omega i}\hat{E}_j$. To find the first of these expectation values, we calculate the frequency-domain correlation function of $\mathbf{\hat{Y}}_\omega$ and $\mathbf{\hat{B}}$. Again we include only terms that depend on the magnetic-field part of $\mathbf{\hat{Y}}_\omega$ in (\ref{YCB}) and denote this fact by a subscript $\scriptstyle B$ on the correlation function. From (\ref{YCB}) and (\ref{BE}) we find 
\begin{eqnarray}
\fl
  \left\langle\mathbf{\hat{Y}}^\dagger_\omega (\mathbf{r}, \omega')\otimes\mathbf{\hat{B}}(\mathbf{r'}, \omega'')\right \rangle_B \nonumber \\[3pt]
\fl
\qquad = -\rmi\frac{\pi}{\omega''}\sqrt{\frac{2\hbar}{\omega}}\delta(\omega-\omega')\left\langle\mathbf{\hat{C}}^\dagger_{\mathrm{m}}(\mathbf{r}, \omega')\otimes\nabla'\times\mathbf{\hat{E}}(\mathbf{r'}, \omega'')\right \rangle  \nonumber \\[3pt]
\fl
\qquad \quad + \frac{\beta (\mathbf{r},\omega)}{2\omega\omega'\omega''} \left[\mathrm{P}\frac{2\omega}{\omega^2-\omega'^2}-\rmi\pi\delta(\omega-\omega')\right]\left\langle\nabla\times\mathbf{\hat{E}}^\dagger(\mathbf{r}, \omega')\otimes\nabla'\times\mathbf{\hat{E}}(\mathbf{r'}, \omega'')\right\rangle.  \label{YBd0}
\end{eqnarray}
The second correlation function on the right-hand side of (\ref{YBd0}) is found from (\ref{EdE}); the first correlation function is shown by (\ref{CdC})--(\ref{CC}), (\ref{EopG}) and (\ref{jopdef}) to be
\begin{eqnarray}
\fl
\left\langle\hat{C}^\dagger_{\mathrm{m}i} (\mathbf{r}, \omega')(\nabla'\times\mathbf{\hat{E}})_j(\mathbf{r'}, \omega'')\right\rangle \nonumber \\[3pt]
\fl
\qquad =2\pi\rmi\mu_0\omega''\left[-\frac{\hbar\kappa_0}{\pi}\kappa_\mathrm{I}(\mathbf{r},\omega'')\right]^{1/2}\mathcal{N}(\omega')(\nabla'\times\mathbf{G}(\mathbf{r'},\mathbf{r},\omega')\times\stackrel{\leftarrow}{\nabla})_{ji}\delta(\omega'-\omega'').  \label{CdcurlE} 
\end{eqnarray}
Insertion of (\ref{EdE}) and (\ref{CdcurlE}) in (\ref{YBd0}), with use of (\ref{ab}), gives
\begin{eqnarray}
\fl
\left\langle\hat{Y}^\dagger_{\omega i}(\mathbf{r}, \omega')\hat{B}_j(\mathbf{r'}, \omega'')\right \rangle_B    \nonumber \\[3pt]
\fl
\quad =2\pi^2\mu_0\sqrt{\frac{2\hbar}{\omega}}\left[-\frac{\hbar\kappa_0}{\pi}\kappa_\mathrm{I}(\mathbf{r},\omega'')\right]^{1/2}\mathcal{N}(\omega')(\nabla'\times\mathbf{G}(\mathbf{r'},\mathbf{r},\omega')\times\stackrel{\leftarrow}{\nabla})_{ji}\delta(\omega-\omega')\delta(\omega'-\omega'')   \nonumber \\[3pt]
\fl
\qquad +\frac{2\pi\hbar\mu_0\omega'}{\omega\omega''}\left[-\frac{2\kappa_0}{\pi}\omega\kappa_\mathrm{I}(\mathbf{r},\omega'')\right]^{1/2} \delta(\omega'-\omega'')\mathcal{N}(\omega') \nonumber \\[3pt]
 \times\left[\mathrm{P}\frac{2\omega}{\omega^2-\omega'^2}-\rmi\pi\delta(\omega-\omega')\right](\nabla\times\mathbf{G}_{\mathrm{I}}(\mathbf{r},\mathbf{r'},\omega')\times\stackrel{\leftarrow}{\nabla'})_{ij}.     \label{YdB}
\end{eqnarray}
It is straightforward to show that the correlation function $\left\langle\hat{Y}_{\omega i}(\mathbf{r}, \omega')\hat{B}^\dagger_j(\mathbf{r'}, \omega'')\right \rangle_B$ differs from (\ref{YdB}) by a complex conjugation and the replacement of $\mathcal{N}(\omega')$ by $\mathcal{N}(\omega')+1$; we can then compute the equal-time correlation function of $\mathbf{\hat{Y}}_\omega$ and $\mathbf{\hat{B}}$ using these frequency-domain results and the general relation (\ref{EEEEf}). The expectation value of interest is $\langle\int_0^\infty\rmd\omega\,\beta (\mathbf{r},\omega)\hat{Y}_{\omega i}\hat{B}_j\rangle_B$; with use of (\ref{ab}) and the Kramer-Kronig relation (\ref{KK}), this  expectation value is found to be 
\begin{eqnarray}
\fl
\left\langle\int_0^\infty\rmd\omega\,\beta (\mathbf{r},\omega)\hat{Y}_{\omega i} (\mathbf{r}, t)\hat{B}_j(\mathbf{r}, t)\right \rangle_B  \nonumber \\[3pt]
 = -\frac{\hbar}{\pi}\mathrm{Im}\int_0^\infty\rmd\omega'\, \left[\kappa (\mathbf{r},\omega')-1\right] \coth\left(\frac{\hbar\omega'}{2k_B T}\right)\Delta^{\!B}_{\ \,ij}(\mathbf{r},\omega').   \label{betaYB}  
\end{eqnarray}
The final expectation value required for our purposes is that of $\int_0^\infty\rmd\omega\,\alpha (\mathbf{r},\omega)\hat{X}_{\omega i}\hat{E}_j$. Only terms that depend on the electric-field part of $\mathbf{\hat{X}}_\omega$ in (\ref{XCE}) are included; this fact is denoted by a subscript $\scriptstyle E$ on the correlation function. The calculation exactly parallels that leading to  (\ref{betaYB}) and the result is
\begin{eqnarray}
\fl
\left\langle\int_0^\infty\rmd\omega\,\alpha (\mathbf{r},\omega)\hat{X}_{\omega i} (\mathbf{r}, t)\hat{E}_j(\mathbf{r}, t)\right \rangle_E  \nonumber \\[3pt]
 = \frac{\hbar}{\pi c^2}\mathrm{Im}\int_0^\infty\rmd\omega'\,\left[\varepsilon (\mathbf{r},\omega')-1\right] \coth\left(\frac{\hbar\omega'}{2k_B T}\right)\Delta^{\!E}_{\ \,ij}(\mathbf{r},\omega').  \label{alphaXE}  
\end{eqnarray}
This completes the set of expectation values needed to compute the electromagnetic part of the energy density and stress tensor in thermal equilibrium. 

\section{Casimir energy density}  \label{sec:en}
In the previous section we ignored the free-field parts of the operators $\mathbf{\hat{X}}_\omega$ and $\mathbf{\hat{Y}}_\omega$. The rationale for this omission in deriving the Casimir effect is that the stress-energy associated with the free-field part of the reservoir represents the absorbed energy due to the dissipation of the medium, together with the zero-point energy of the reservoir. The dissipated energy is included in the canonical theory of macroscopic electromagnetism, with the result that the system is closed and a proper quantization can be performed~\cite{phi10}. The description as a closed system is also essential to the existence of a stress-energy-momentum tensor, derived in section~\ref{sec:em}. But Casimir forces are caused by the stress-energy of the electromagnetic fields, not by the stress-energy absorbed and dissipated in the medium. This is why the correlation functions of the previous section, which include only the electromagnetic contribution, determine the Casimir energy density and stress.

The classical expression (\ref{rho}) for the energy density also holds in the quantum theory, except that the final term must be written in a hermitian form to give a hermitian energy-density operator. We thus have the operator
\begin{eqnarray}
\fl
\hat{\rho}=&\frac{\kappa_0}{2}\left[\frac{1}{c^2}\mathbf{\hat{E}}^2+\mathbf{\hat{B}}^2\right]  \nonumber \\[3pt]
\fl
&+\int_0^\infty\rmd\omega\left[\frac{1}{2}(\partial_t\mathbf{\hat{X}}_\omega)^2+\frac{1}{2}(\partial_t\mathbf{\hat{Y}}_\omega)^2+\frac{1}{2}\omega^2(\mathbf{\hat{X}}_\omega^2+\mathbf{\hat{Y}}_\omega^2)-\frac{1}{2}\beta\left(\mathbf{\hat{Y}}_\omega\cdot\mathbf{\hat{B}}+\mathbf{\hat{B}} \cdot\mathbf{\hat{Y}}_\omega\right)\right]. \label{rhoop}
\end{eqnarray}
The $\beta$-dependent term in (\ref{rhoop}) has an expectation value that follows immediately from (\ref{betaYB}); the hermitian combination in (\ref{rhoop}) just picks out the real part of (\ref{betaYB}), which is the entire right-hand side. The expectation value of the quadratic terms in the electric and magnetic fields in (\ref{rhoop}) are obtained from (\ref{EE}) and (\ref{BB}). When all terms are combined, the final expression for the Casimir energy density $\langle\hat{\rho}\rangle$ is
\begin{eqnarray}
\fl
\left\langle\hat{\rho}\right\rangle =\frac{\hbar}{2\pi}\mathrm{Im}\int_0^\infty\rmd\omega \, \coth\left(\frac{\hbar\omega}{2k_B T}\right)&\left\{ \frac{1}{c^2}\frac{\rmd[\omega\varepsilon(\mathbf{r},\omega)]}{\rmd\omega} \Delta^{\!E\ i}_{\ \,i}(\mathbf{r},\omega)\right.  \nonumber \\[3pt]
&\left. \ \, +\left[\kappa(\mathbf{r},\omega)-\omega\frac{\rmd\kappa(\mathbf{r},\omega)}{\rmd\omega}\right] \Delta^{\!B\ i}_{\ \,i}(\mathbf{r},\omega)\right\}.      \label{rhocas}
\end{eqnarray}
Note that the $\kappa$-dependent factor in (\ref{rhocas}) takes the form $[\rmd(\omega\mu)/\rmd\omega]/\mu^2$ when written in terms of $\mu=1/\kappa$. Casimir forces at zero temperature can be found by using (\ref{rhocas}) to calculate the total Casimir energy of a configuration of objects and taking derivatives with respect to the parameters specifying their separations and relative orientations.

The factors in (\ref{rhocas}) that depend on the dielectric functions have a familiar form: the Brillouin expression~\cite{jac,LLcm} for the monochromatic electromagnetic energy density of a {\it lossless} medium contains the same quantities, where $\varepsilon$ and $\kappa$ are real in that case. The result (\ref{rhocas}) is in many ways remarkably simple, considering that it holds for arbitrary dispersion (consistent with the Kramers-Kronig relations). Dispersion has a highly complicating effect on the electromagnetic energy of general fields in media, even when the difficulty of losses can be ignored~\cite{phi11}. Here the losses are compensated because of the imposition of thermal equilibrium, and the restriction to thermal (and zero-point) fields also has the special effect that dispersion contributes only through the simple first-order frequency derivatives in (\ref{rhocas}). 

For computational purposes it is more convenient to re-express the frequency integral in (\ref{rhocas}) as a sum over imaginary frequencies. Because of the property~\cite{LL}
\begin{equation}  \label{GG*}
\mathbf{G}(\mathbf{r},\mathbf{r'},-\omega)=\mathbf{G}^*(\mathbf{r},\mathbf{r'},\omega)
\end{equation}
of the Green bi-tensor, its real part is even in $\omega$ while is imaginary part is odd, and the same property holds for $ \Delta^{\!E}_{\ \,ij}(\mathbf{r},\omega)$ and $ \Delta^{\!B}_{\ \,ij}(\mathbf{r},\omega)$ (recall (\ref{DeltaE}) and (\ref{DeltaB})). Moreover, the dielectric functions also have the property (\ref{GG*})~\cite{LL}. The hyperbolic cotangent in (\ref{rhocas}), on the other hand, is odd in $\omega$. All this means that the imaginary part of the integral in (\ref{rhocas}) is automatically extracted if we modify the integration over $\omega$ so that it runs from $-\infty$ to $\infty$ and multiply by by $-\rmi/2$; thus  (\ref{rhocas}) can be replaced by
\begin{eqnarray}
\fl
\left\langle\hat{\rho}\right\rangle =-\rmi\frac{\hbar}{4\pi}\int_{-\infty}^\infty\rmd\omega \, \coth\left(\frac{\hbar\omega}{2k_B T}\right)&\left\{ \frac{1}{c^2}\frac{\rmd[\omega\varepsilon(\mathbf{r},\omega)]}{\rmd\omega} \Delta^{\!E\ i}_{\ \,i}(\mathbf{r},\omega)\right.  \nonumber \\[3pt]
&\left. \ \, +\left[\kappa(\mathbf{r},\omega)-\omega\frac{\rmd\kappa(\mathbf{r},\omega)}{\rmd\omega}\right] \Delta^{\!B\ i}_{\ \,i}(\mathbf{r},\omega)\right\}.      \label{rhocas2}
\end{eqnarray}
We can now close the frequency integral in the upper-half complex frequency plane, where the dielectric functions and the Green bi-tensor are analytic~\cite{LLcm,LL}. This contour integral is given by the sum of the residue contributions from the poles in the hyperbolic cotangent term at positive imaginary frequencies $\omega=2\pi\rmi k_BTn/\hbar$, $n=0,1,2,\dots$. With the notation
\begin{equation}  \label{xi}
\rmi \xi_n:=\rmi \frac{2\pi k_BTn}{\hbar}, \qquad n=0,1,2,\dots,
\end{equation}
for these imaginary frequencies, the Casimir energy density (\ref{rhocas2}) then has the form
\begin{eqnarray}
\left\langle\hat{\rho}\right\rangle =k_BT{\sum_{n=0}}'&\left\{ \frac{1}{c^2}\left.\frac{\rmd[\omega\varepsilon(\mathbf{r},\omega)]}{\rmd\omega}\right|_{\omega=\rmi\xi_n} \Delta^{\!E\ i}_{\ \,i}(\mathbf{r},\rmi\xi_n)\right.  \nonumber \\[3pt]
&\left. \ \, +\left[\frac{1}{\mu^2(\mathbf{r},\omega)}\frac{\rmd[\omega\mu(\mathbf{r},\omega)]}{\rmd\omega}\right]_{\omega=\rmi\xi_n} \Delta^{\!B\ i}_{\ \,i}(\mathbf{r},\rmi\xi_n)\right\},      \label{rhocasim}
\end{eqnarray}
where the prime on the summation sign means that the first term in the sum is taken with a factor of $1/2$ (because the contour passes through the pole at $\omega=0)$, and where $\kappa$ has been replaced by $\mu(=1/\kappa)$. At zero temperature, where only the zero-point contribution remains, the sum in (\ref{rhocasim}) becomes an integral over positive imaginary frequencies:
\begin{equation}
\fl
\left\langle\hat{\rho}\right\rangle_{T=0} =\frac{\hbar}{2\pi}\int_0^\infty\rmd\xi\left\{ \frac{1}{c^2}\frac{\rmd[\xi\varepsilon(\mathbf{r},\rmi\xi)]}{\rmd\xi} \Delta^{\!E\ i}_{\ \,i}(\mathbf{r},\rmi\xi)
+\frac{1}{\mu^2(\mathbf{r},\rmi\xi)}\frac{\rmd[\xi\mu(\mathbf{r},\rmi\xi)]}{\rmd\xi} \Delta^{\!B\ i}_{\ \,i}(\mathbf{r},\rmi\xi)\right\}.      \label{rhocasim0}
\end{equation}

The correct expression for the Casimir energy density in media has been a subject of conflicting assertions; as pointed out in~\cite{phi10b}, however, only the form (\ref{rhocasim0}) gives a Casimir force between parallel plates that agrees with the force obtained from the vacuum stress tensor between the plates. The correct form of the energy density emerges automatically from macroscopic QED, which is moreover a fully quantum treatment of the problem.

\section{Casimir stress tensor}  \label{sec:st}
The quantum stress tensor operator has the same form as the classical expression (\ref{stress}), when the latter is written so as to give a hermitian operator: 
\begin{eqnarray}
\fl
\hat{\sigma}_{ij}&=\frac{1}{2}\delta_{ij}(\varepsilon_0\mathbf{\hat{E}}^2+\kappa_0\mathbf{\hat{B}}^2)-\varepsilon_0\hat{E}_i\hat{E}_j-\kappa_0\hat{B}_i{B}_j  \nonumber  \\[3pt]
\fl
&+\int_0^\infty\rmd\omega\left\{ \delta_{ij}\left[\frac{1}{2}(\partial_t\mathbf{\hat{X}}_\omega)^2+\frac{1}{2}(\partial_t\mathbf{\hat{Y}}_\omega)^2-\frac{1}{2}\omega^2(\mathbf{{\hat{X}}}_\omega^2+\mathbf{{\hat{Y}}}_\omega^2)+\frac{1}{2}\alpha \left(\mathbf{{\hat{X}}}_\omega\cdot \mathbf{{\hat{E}}}+\mathbf{{\hat{E}}} \cdot\mathbf{{\hat{X}}}_\omega\right)\right]\right.  \nonumber  \\[3pt]
\fl
& \qquad \qquad \quad -\frac{1}{2}\alpha\left( \hat{E}_i\hat{X}_{\omega j}+\hat{X}_{\omega j}\hat{E}_i\right)+\frac{1}{2}\beta \left(\hat{Y}_{\omega i}\hat{B}_j+\hat{B}_j \hat{Y}_{\omega i}\right){\Bigg\}}. \label{stressop}
\end{eqnarray}
The Casimir stress tensor is the expectation value of the electromagnetic part of (\ref{stressop}) in thermal equilibrium. We proceed as in the case of the energy density in the last section.

The $\alpha$- and $\beta$-dependent terms in (\ref{stressop}) have expectation values that follow directly from (\ref{alphaXE}) and (\ref{betaYB}), and the expectation values of the terms quadratic in the electric and magnetic fields are obtained from (\ref{EE}) and (\ref{BB}). The Casimir stress tensor $\left\langle\hat{\sigma}_{ij}\right\rangle$ is thereby found to be
\begin{eqnarray}
\fl
\left\langle\hat{\sigma}_{ij}\right\rangle =\frac{\hbar}{\pi}\mathrm{Im}\int_0^\infty\rmd\omega \, \coth\left(\frac{\hbar\omega}{2k_B T}\right)&\left\{ \frac{1}{c^2}\varepsilon(\mathbf{r},\omega)\left[\frac{1}{2}\delta_{ij}\Delta^{\!E\ k}_{\ \,k}(\mathbf{r},\omega)-\Delta^{\!E}_{\ \,ij} (\mathbf{r},\omega)\right]\right.  \nonumber \\[3pt]
&\left. \ \, +\kappa(\mathbf{r},\omega)\left[\frac{1}{2}\delta_{ij}\Delta^{\!B\ k}_{\ \,k}(\mathbf{r},\omega)-\Delta^{\!B}_{\ \,ij} (\mathbf{r},\omega)\right] \right\}.      \label{stresscas}
\end{eqnarray}
As in the case of the Casimir energy density, a computationally more convenient formula for the stress tensor involves a sum over imaginary frequencies. The derivation is as described in leading up to (\ref{rhocasim}) and the expression is
\begin{eqnarray}
\left\langle\hat{\sigma}_{ij}\right\rangle =2k_BT{\sum_{n=0}}'&\left\{ \frac{1}{c^2}\varepsilon(\mathbf{r},\rmi\xi_n)\left[\frac{1}{2}\delta_{ij}\Delta^{\!E\ k}_{\ \,k}(\mathbf{r},\rmi\xi_n)-\Delta^{\!E}_{\ \,ij} (\mathbf{r},\rmi\xi_n)\right]\right.  \nonumber \\[3pt]
&\left. \ \, +\kappa(\mathbf{r},\rmi\xi_n)\left[\frac{1}{2}\delta_{ij}\Delta^{\!B\ k}_{\ \,k}(\mathbf{r},\rmi\xi_n)-\Delta^{\!B}_{\ \,ij} (\mathbf{r},\rmi\xi_n)\right] \right\}.      \label{stresscasim}
\end{eqnarray}
At zero-temperature we obtain from (\ref{stresscasim}) the zero-point Casimir stress as an integral over positive imaginary frequencies:
\begin{eqnarray}
\left\langle\hat{\sigma}_{ij}\right\rangle =\frac{\hbar}{\pi}\int_0^\infty\rmd\xi&\left\{ \frac{1}{c^2}\varepsilon(\mathbf{r},\rmi\xi)\left[\frac{1}{2}\delta_{ij}\Delta^{\!E\ k}_{\ \,k}(\mathbf{r},\rmi\xi)-\Delta^{\!E}_{\ \,ij} (\mathbf{r},\rmi\xi)\right]\right. \nonumber \\[3pt]
&\left. \ \, +\kappa(\mathbf{r},\rmi\xi)\left[\frac{1}{2}\delta_{ij}\Delta^{\!B\ k}_{\ \,k}(\mathbf{r},\rmi\xi)-\Delta^{\!B}_{\ \,ij} (\mathbf{r},\rmi\xi)\right] \right\}.      \label{stresscasim0}
\end{eqnarray}

The formula (\ref{stresscasim}) (with $\kappa=1$) for the Casimir stress tensor in media was obtained by Herculean efforts in~\cite{dzy61}, and is the most general result of Lifshitz theory. Part of the reason why such an enormously complicated formalism was required in~\cite{dzy61} is the lack of a Hamiltonian and Lagrangian basis for the theory, which also undermines any claims that the result applies to quantum electromagnetic fields (see Introduction). As in the case of the energy density in the last section, the Casimir stress tensor in media emerges  in a self-contained manner from macroscopic QED in thermal equilibrium, without the need for additional input.

It is interesting also to compute the expectation value in thermal equilibrium of the quantum version of (\ref{conminhom}). The first term on the left-hand side of (\ref{conminhom}) has of course zero expectation value in the stationary situation of thermal equilibrium. The right-hand side must be written in hermitian form in the quantum theory, and using $\alpha\nabla_i\alpha=\nabla_i(\alpha^2)/2$ and $\beta\nabla_i\beta=\nabla_i(\beta^2)/2$ its expectation value is found in a manner similar to that used to obtain (\ref{alphaXE}) and (\ref{betaYB}). This leads to the following result for the divergence of the Casimir stress tensor:
\begin{eqnarray}
\fl
\left\langle\nabla_j\hat{\sigma}_{i}^{\ j}\right\rangle &=\frac{\hbar}{2\pi}\mathrm{Im}\int_0^\infty\rmd\omega \, \coth\left(\frac{\hbar\omega}{2k_B T}\right)\left[ \frac{1}{c^2}\Delta^{\!E\ j}_{\ \,j}(\mathbf{r},\omega)\nabla_i \varepsilon(\mathbf{r},\omega)\right.    \nonumber  \\[3pt]
\fl
 & \qquad \qquad \qquad \qquad\qquad \qquad\   -\Delta^{\!B\ j}_{\ \,j}(\mathbf{r},\omega)\nabla_i\kappa(\mathbf{r},\omega)\Bigg]      \label{divstresscas}  \\
 \fl
&= k_BT{\sum_{n=0}}'\left[ \frac{1}{c^2}\Delta^{\!E\ j}_{\ \,j}(\mathbf{r},\rmi\xi_n)\nabla_i \varepsilon(\mathbf{r},\rmi\xi_n)-\Delta^{\!B\ j}_{\ \,j}(\mathbf{r},\rmi\xi_n)\nabla_i\kappa(\mathbf{r},\rmi\xi_n)\right].
\end{eqnarray}

As a final remark on the Casimir stress-energy-momentum tensor, we note that the energy flux and momentum density in this tensor must be zero, because the electromagnetic fields are in thermal equilibrium. It is straightforward to verify, by calculations similar to those used to obtain (\ref{rhocas}) and (\ref{stresscas}), that the electromagnetic parts (defined as in section~\ref{sec:cor}) of the energy flux (\ref{s}) and momentum density (\ref{p}) do indeed vanish in thermal equilibrium.

\section{Conclusions}
The Casimir effect has been derived from macroscopic QED~\cite{phi10} by a simple restriction to thermal equilibrium. Expressions for the Casimir energy density and stress tensor were obtained for arbitrary inhomogeneous magnetodielectrics.  As the results are derived from a rigorous quantization of electromagnetic fields in dispersive, dissipative media, they not subject to the criticisms that have been directed at the standard Lifshitz theory of the Casimir effect. Moreover, the canonical basis of macroscopic QED~\cite{phi10} means that the correct forms of the Casimir energy density and stress tensor in media emerge directly from the theory. In Lifshitz theory, by contrast, there is no Hamiltonian or Lagrangian, so that detailed mechanical and thermodynamical arguments are required to obtain the form of the electromagnetic stress tensor in media.

\ack
This research is supported by the Scottish Government and the Royal Society of Edinburgh.

\end{document}